\documentclass[a4paper,10pt]{article}
 
\usepackage{amssymb}
\usepackage{amsthm}
\usepackage{amsmath}
\usepackage[ruled,linesnumbered]{algorithm2e}
\usepackage{graphicx}
\usepackage{epstopdf}
\usepackage{url} 
\usepackage{fancyhdr}
\usepackage{subcaption}
\usepackage{multirow}
\usepackage{color}
\usepackage{textcomp}

\title{Towards a Standard Sampling Methodology on Online Social Networks: Collecting Global Trends on Twitter}
\author{
        C. A. Pi\~na-Garc\'ia \\
				Instituto de Investigaciones en Matem\'aticas Aplicadas y en Sistemas\\
         Departamento de Ciencias de la Computaci\'on\\				 
         Universidad Nacional Aut\'onoma de M\'exico, Distrito Federal, M\'exico\\
				\textbf{carlos.pgarcia@iimas.unam.mx}
				\and
        Dongbing Gu\\
        School of Computer Science and Electronic Engineering\\
         University of Essex\\
        Wivenhoe Park, Colchester, UK\\
				\textbf{dgu@essex.ac.uk}
				}

\begin{document}

\maketitle

\begin{abstract}
One of the most significant current challenges in large-scale online social networks, is to establish a concise and coherent method able to collect and summarize data. Sampling the content of an Online Social Network (OSN) plays an important role as a knowledge discovery tool. 

It is becoming increasingly difficult to ignore the fact that current sampling methods must cope with a lack of a full sampling frame i.e., there is an imposed condition determined by a limited data access. In addition, another key aspect to take into account is the huge amount of data generated by users of social networking services. This type of conditions make especially difficult to develop sampling methods to collect truly reliable data. Therefore, we propose a low computational cost method for sampling emerging global trends on social networking services such as Twitter. 
  
The main purpose of this study, is to develop a methodology able to carry out an efficient collecting process via three random generators: Brownian, Illusion and Reservoir. These random generators will be combined with a Metropolis-Hastings Random Walk (MHRW) in order to improve the sampling process. We demonstrate the effectiveness of our approach by correctly providing a descriptive statistics of the collected data. In addition, we also sketch the collecting procedure on real-time carried out on Twitter. Finally, we conclude with a trend concentration graphical description and a formal convergence analysis to evaluate whether the sample of draws has attained an equilibrium state to get a rough estimate of the sample quality. 

\end{abstract}

\section{Introduction}

In recent years, there has been an increasing interest in Online Social Networks (OSNs) exploration. This type of research has become a major and necessary area in terms of data collection. More recently,  areas of social networking analysis are now expanding to different disciplines, not only in data mining studies but also in computational social science (user behavior), social media analytics, marketing applications and discovery with social Big Data. Researchers in these areas have highlighted the potential of using big data information to transform and enhance businesses (e.g.,knowledge management and decision making). In addition, the availability of unprecedented amounts of data about human interactions from different social networks opens the possibility of using those data to leverage knowledge about social behavior and the activity of individuals \cite{weng2012competition}.

Mobile phone data and the content generated by hundreds of millions of users on social media such as: Twitter or Facebook presents continuous data streams of human social activities, and offer an unprecedented opportunity to understand the structure and dynamics of social behavior in various situations. Thus, it would be fair to say that people spend an increasing amount of time every day consuming information. 

This research seeks to present a new methodology to produce samples of ``trending topics'' generated in real-time by Twitter users. Twitter is the most famous microblogging website in the social media space, where users post messages that are limited to 140 characters. In addition, users can follow others they find interesting, the posts are called ``tweets''. Unlike the case with many other social networks, the relationship does not have to be mutual \cite{golbeck2013analyzing}. It should be noted that Twitter produces approximately 500 million tweets per day, with 200 million of regular users. Therefore, Twitter has been a valuable tool to track and to identify patterns of mobility and activity, especially using geolocated tweets. Geolocated tweets typically use the Global Positioning System (GPS) tracking capability installed on mobile devices when enabled by the user to give his or her precise location (latitude and longitude). In this regard, it is fair to assume that Twitter represents a suitable large-scale social network to be explored. 

Owing to the size of the entire social network, which is problematic to be measured, the analysis of the entire network is infeasible and sampling is unavoidable. In this case, random sampling offers an efficient tool for collecting reliable and unbiased data. Most of the samples are analyzed, and results are generalized to the population from which the samples were drawn \cite{maiya2010sampling}.  

This study shows that researchers need to understand and exploit data in order to uncover patterns, relationships and trends from real-world phenomena. In particular, it is observed and quantified the patterns that emerge naturally from different social networks. A central hypothesis in this work is that in order to advance our understanding of social interaction, it is necessary to propose a new breed of social intelligence platforms able to collect, analyze and visualize information.

\subsection{Contributions}

It becomes necessary to obtain representative data. Therefore, the aim of this study is to propose an algorithm to discover and collect emerging global trends on Twitter. Specifically, our contributions in this study  are as follows:

\begin{itemize}

	\item This paper provides a series of random generators (Brownian, Illusion and Reservoir) based on random walk models to sample small but relevant parts of information produced on Twitter. In addition, this research is intended to determine the extent to which random walks can be combined by using an alternative version of a Metropolis-Hastings algorithm. We can consider this technique as an efficient real-time tool for summarizing OSNs and quickly make sense of big datasets. This paper seeks to collect and examine a set of samples of global trends\footnote{A word, phrase or topic that is tagged at a greater rate than other tags is said to be a trending topic.} provided by Twitter.
	
	\item This study compares the performance of three random generators. In this regard, it can be argued that on-line sampling techniques could benefit from integrating efficient and rapid methods to find relevant trending content on Twitter. 
	
	\item We outline research challenges and opportunities in leveraging social networks for future social media study. In addition, this approach therefore provides an important opportunity to advance the understanding of using different sampling methods that may help to set up strategic insight and recommendations via a range of customized sampling methods.
\end{itemize}

The outline of the rest of this paper is as follows: in Section \ref{sec:rel} we explain the current work and a literature review related to our study. Similarly, in Section \ref{sec:MH} gives a brief overview of the Metropolis-Hasting Random Walk algorithm. Section \ref{sec:RA} defines briefly a graph notation and presents our three random generators ($q(y|x)$: Brownian, Illusion and Reservoir) that we intend to incorporate into the alternative MHRW. Section \ref{sec:algo} explains the methodology to build an alternative Metropolis-Hasting Random Walk and how this technique can be applied to generate a candidate node via our three random generators. In Section \ref{sec:sampling} we explain how to collect global trending topics from Twitter via its API. In Section \ref{sec:results} we evaluate and compare the performance of our implementation in terms of the amount of sampled trends, number of followers per trend, duplicate trends and memory usage. Additionally, we also report the convergence analysis to get a rough estimate of the sample quality. Section \ref{sec:limit} is concerned with the limitations of this research. Finally, Sections \ref{sec:con} and \ref{sec:future} concludes the paper and provides a future work plan.

\section{Related Work}\label{sec:rel}

A considerable amount of literature has been published on using graph sampling techniques on large-scale OSNs. These studies are rapidly growing in the scientific community, showing that sampling methods are essential for practical estimation of OSN properties. These properties include, for example: user age distribution, net activity, net connectivity and node degree. Studies on social science show the importance of graph sampling techniques, e.g., \cite{scott2011social,lee2006statistical,caci2012facebook,fire2013organization,mislove2006exploiting}. 

Online social networks such as Facebook represents one of the biggest social services in the world. Therefore, it may be seen as a large-scale source to collect data with the aim to obtain a representative sample or characterize the whole network structure \cite{ugander2011anatomy,bhattacharyya2011analysis,ferri2012new,caci2011facebook}. Recent evidence suggests that efficient random walk inspired techniques has been successfully used to sample large-scale social networks, in particular, Facebook \cite{gjoka10_walkingfb,gjoka2011practical}. However, despite its relative success of Facebook, these specific sampling strategies have not been tested on different social networking services such as: Twitter.   

A number of researchers have pointed out that statistical approaches such as random walks can be used to improve and speeding up the process of sampling. This can be done by considering different randomized algorithms which are able to cope with large datasets. Recently, the Metropolis-Hastings Random Walk algorithm have been tested on Facebook and \texttt{Last.fm} (a music website with 30 million active users) showing significant results for an unbiased sampling of users \cite{gjoka2011practical,kurant11_magnifying,gjoka2011multigraph}. 

Similarly, some studies based on supervised random walks use the information from the network structure with the aim to guide a random walk on the graph, e.g., on the Facebook social graph \cite{backstrom2011supervised}. In addition, there are other studies that introduce the same random walk technique analyzed from the Markov Chain Monte Carlo (MCMC) perspective, i.e., the Metropolis-Hastings random walk (MHRW), which is mainly used to produce uniform samples \cite{bar2008random}. 

An alternative Metropolis-Hastings random walk using a spiral proposal distribution is presented in \cite{pina2013spiraling}. The authors examined whether it was possible to alter the behavior of the MHRW using spirals as a probability distribution instead a classic \textit{Gaussian} distribution. They observed that the spiral inspired approach was able to adapt itself correctly to a Metropolis-Hastings random walk.
 
These studies presented thus far provide evidence that there is a growing interest in the use of rapid sampling models and a clear need of data extraction tools on Facebook \cite{ugander2011anatomy,bhattacharyya2011analysis,ferri2012new,caci2011facebook}. However,  Twitter has recently received special attention from researchers that are interested in uncovering global topics, that are well known as: ``memes'' and ``hashtags\footnote{A hashtag is a word or  metadata tag prefixed with the hash symbol (\#).}'' \cite{hawelka2013geo,takhteyev2012geography,mitchell2013geography,thapen2013towards,kallus2014predicting}.

Recently, there has been an increasing amount of literature on data collection via Twitter. Preliminary work on information diffusion was presented in \cite{weng2013role}, where authors examined the mechanisms behind human interactions through an unprecedented amount of data (social observatory). They also argued that information diffusion affects network evolution.

An important analysis about the geography of twitter networks was presented in \cite{takhteyev2012geography}. In this case, the authors showed that distance matters on Twitter, both at short and longer ranges. In addition, they argued that the distance considerably constrains ties. The authors highlighted the importance of Twitter in terms of collection of data due to its popularity and international reach. They also suggested that these ties at distances of up to 1000 km are more frequent than what it would be expected if the ties were formed randomly.

In a large longitudinal study carried out in \cite{hawelka2013geo}, the authors found global patterns of human mobility based on data extracted from Twitter. A dataset of almost a billion of tweets recorded in 2012, was used to estimate volumes of international travelers. The authors argue that Twitter is a viable source to understand and quantify global mobility patterns.  

Furthermore, a detailed investigation on correlations between real-time expressions of individuals and a wide range of emotional, geographic, demographic and health characteristics was conducted in \cite{mitchell2013geography}. Results showed how social media may potentially be used to estimate real-time levels and changes in population-level measures. The findings in \cite{mitchell2013geography}, were supported by a large dataset of over 10 million geo-tagged tweets, gathered from 373 urban areas in the United States during the calendar year of 2011.

In another major study, a ``conversational vibrancy'' framework to capture dynamics of hashtags based on their topicality, interactivity, diversity, and prominence was introduced in \cite{lin2013bigbirds}. The authors examined the growth and persistence of hashtags during the 2012 U.S. presidential debates. They point out that the growth of a hashtag and death is largely determined by an environmental context condition rather than the conversational vibrancy of the hashtag itself.   

A recent study about community structure in OSNs was presented in \cite{Lilian2013srep}. Authors claim that ``memes'' and behaviors can be mimicked as a contagion phenomenon. The authors concluded that the future popularity of a meme can be predicted by quantifying its early spread patterns.

A very important study that builds a systematic framework for investigating human behaviors under extreme events with online social network data extracted from Twitter was carried out in \cite{lu2014network}. The researchers have shown distinctive changes in patterns of interactions in online communities that have been affected by a natural disaster compared to communities that were not affected.  

Finally, in a controlled study of the automatic analysis of UK political tweets was provided in \cite{thapen2013towards}. In this case, the authors examined the extent of which the volume and sentiment of tweets can be used as a proxy to obtain their voting intentions and compare the results against existing poll data. In addition, the authors propose a data collection method through a list of selected Twitter accounts classified by party affiliation. Approximately 689,637 tweets, were retrieved from the publicly available \texttt{timelines} of the members of Parliament on 10th June 2013, the authors took a random sample of 600 users from Twitter.

\section{The Metropolis-Hastings Algorithm} \label{sec:MH}

Metropolis-Hastings algorithms are a major area of interest within the field of Markov chain Monte Carlo (MCMC) methods. In addition, this algorithm was selected as one of the top 10 algorithms of the twentieth century \cite{cipra2000best}. The Metropolis׈astings algorithm can be viewed as generic Markov chain Monte Carlo algorithm \cite{martinez2001computational,robert2009introducing}. This method obtains the state of the chain at $t+1$ by sampling a candidate point $Y$ from a proposal distribution $q$. The candidate point is accepted as the next state of the chain with probability given by          

\begin{equation}\label{ch5eq:point}
\rho(x,y) = \min\left\{\frac{f(y)q(x|y)}{f(x)q(y|x)},1 \right\} \quad . 
\end{equation}  

If the candidate point $Y$ is not accepted, then the chain does not move and $X^{(t+1)} = X^{(t)}$. According to \cite{robert2009introducing}, the Metropolis-Hastings algorithm associated with the objective (target) density \textit{f} and the conditional density $q$ produces a Markov chain $X^{(t)}$ through the following transition kernel:

\begin{enumerate}	
	\item Generate a candidate point $Y_t$ from a given proposal distribution (usually a normal distribution): $q(y|x^{(t)})$.
	\item Generate $U$ from a uniform $(0,1)$ distribution.
	\item If $U \leq \rho(x,y)$ then set $X^{(t+1)} = Y_t$, else set $X^{(t+1)} = X^{(t)}$.	
\end{enumerate}

$q(y|x)$ is a term frequently used in the literature as a random generator and the probability $\rho(x,y)$ is referred as the Metropolis-Hastings acceptance probability. It should be noted that modifying the proposal  distribution $q$ may have interesting consequences on this algorithm.

\section{Defining a Graph Notation and the Random Generators}\label{sec:RA} 

Before defining our proposed random generators, it is important to describe the specific notation that most of the authors employ in their studies, and we will apply for the rest of this paper. An OSN can be easily represented by an undirected graph $G$, which in this case it is referred to the Twitter social network, where $G = (V,E)$, being $V = \left\{v: v = vertices\right\}$ where $v$ stands for a node, the two terms: nodes or vertices can be used interchangeably. Similarly, $E = \left\{e: e = edges \right\}$ where $e$  refers to an edge or a link. An edge indicates a relationship in the network structure. Thus, edges can be either directed or undirected. An undirected edge indicates a mutual relationship, whereas a directed edge indicates a relationship that one node has with the other that is not necessarily reciprocated \cite{golbeck2013analyzing}. In this context, global trends on Twitter are conceived as single nodes that belongs to a graph $G$, which essentially is a group of nodes connected by links or edges.

This section will examine three random generators that are incorporated into the  alternative version of the MHRW. These random generators are aimed to be used heuristically as an internal picker for a candidate node (hereinafter referred to as $\varrho$). This set of random strategies is composed as follows: \textit{Brownian} walk (normal distribution), a spiral-inspired walk (Illusion) and a Reservoir sampling method.  It is important to note that the Brownian case will be used as a the baseline to be compared with the rest of the random generators.

According to Section \ref{sec:MH}, the main idea of the Metropolis-Hastings algorithm is to provide a number of random samples from a given distribution. Thus, our proposed version of the MHRW is able to sample a candidate node $\varrho$, which is directly obtained from: $q(y|x)=$(Brownian, Illusion, Reservoir).

\subsection{Brownian Walk}

The traditional approach used  to sample through the MHRW is based on the normal distribution. This distribution appears in nature due to the wide applicability of the central limit theorem, which states that the distribution for the sum of a large number of statistically independent and identically distributed random variables that have a finite variance converges to a Gaussian \cite{viswanathan2011physics}. In this regard, we have developed a \textit{Brownian} walk that presents a normal distribution, it was decided to use a Brownian walk as a conceptual model of reference.

This model presents a probability distribution for total distance covered in a random walk (biased or unbiased) that tends toward a normal distribution (central limit theorem \cite{dudewicz1976introduction}). Thus, a Brownian walk consists of a series of steps (possibly of different sizes) in randomly chosen directions \cite{codling2010diffusion}. It is important to note that in most cases the Brownian walk is related to a continuous time process. However, in this research it has been considered a discretized version of this strategy. Technically speaking in this model, the candidate node $\varrho$ will be computed according to the Java language command: \texttt{Math.random()}. 

This type of random walk presents explorations over short distances, which can be made in much shorter times than explorations over long distances \cite{berg1993random}. In addition, the random walker tends to explore a given region of space and after that, it tends to return to the same point many times before finally wandering away. It can be said that the random walker chooses new regions to explore blindly and it has no tendency to move toward regions that it has not occupied before.

\subsection{Illusion spiral}

In this research, we have considered a spiral-inspired approach in terms of an \textit{Illusion} spiral.\footnote{see \cite{davis1993spirals} for a full description of the Illusion spiral.} This spiral presents an interesting geometric shape which presents a sequence of points spirally on a plane such that they are equitably and economically spaced (see Fig. \ref{fig:illu}). This spiral model is produced by the following expression: 

\begin{equation}\label{eq:illu}
z \leftarrow az + bz/\left|z\right| \quad . 
\end{equation}  

Where $a = 0.6 + 0.8i$ and $b = 0.65 + 0.7599i$. It is important to note that this spiral-inspired approach should not be considered as a formal distribution in itself. It is just a consequence of making use of complex numbers to correctly generate a geometric shape or in this case an Illusion spiral. In addition, equation \ref{eq:illu} includes complex variables to correctly generate the pattern. 

Thus, a $z$ value ($z \in \mathbb{C}$) of the form $z = a + bi$ is iteratively generated. In this regard, we are able to obtain a collection of complex numbers where the real part will be considered as a candidate node $z = \varrho$ for our study purposes. It should be noted that the experimental methodology regarding the Illusion spiral is based on the work carried out by Pi\~na-Garc\'ia and Gu  in \cite{pina2013spiraling}, the authors analyzed the changing nature of the Metropolis-Hasting Random Walk (MHRW) when $q(y|x)$ is modified by a spiral inspired approach.

\subsection{Reservoir sampling}

A reservoir sampling can be seen as an algorithm that consists in selecting a random sample of size $n$, from a file containing $N$ records, in which the value of $N$ is not known to the algorithm. Throughout this section, the term ``file'' refers to the total number of collected nodes.

According to \cite{vitter1985random}, the first step of any reservoir algorithm is to put the first $n$ records of the file into a ``reservoir''. The rest of the records are processed sequentially. Thus, the number of items to select ($k$) is smaller than the size of the source array $S(i)$. Algorithm \ref{ch7:alg:Reserv} provides an overview of the steps carried out by the reservoir sampling process.

\begin{algorithm} 
\LinesNumbered
\SetAlgoNoLine  
 
 array R[k]  \tcp*[l]{result}
 integer i,j\;
 
 \tcc{fill the reservoir array} 
 \ForEach{$i$ in 1 to $k$}{
 R[i] := S[i]\; 
 }
 
 \tcc{replace elements with gradually decreasing probability}
 \ForEach{i in $k+1$ to lenght(S)}{
 j := random(1,i)\; 
 	\If{$j \leq k$}{
 		R[j] := S[i]\; 
 	} 
 }   
 
\caption{Reservoir Algorithm.}
\label{ch7:alg:Reserv}
\end{algorithm} 

Algorithm \ref{ch7:alg:Reserv} creates a ``reservoir'' array of size $k$ and populates it with the first $k$ items of $S$. It then iterates through the remaining elements of $S$ until this is exhausted. At the $i^{th}$ element of $S$, the algorithm generates a random number $j$ between 1 and $i$. If $j$ is less than $k$, the $j^{th}$ element of the reservoir array is replaced with the $i^{th}$ element of $S$.

In this study,  a non-conditional version of the aforementioned algorithm will be considered i.e., lines 8-10 are discarded. In this case, the random number $j := random(1,i)$  has come to be used to refer to a candidate node $j = \varrho$.

In this context we conceive a random generator as a procedure designed to generate a sequence of random numbers, in this research this sequence of random numbers represents global trends on Twitter. One of the three random generators is selected to pick different trends from list $\mathbf{A}$. Subsequently, all the trends that were sampled are copied to a matrix $\mathbf{B}$, these collected trends are arranged according to how they were chosen.

\section{The Alternative Version of the Metropolis-Hastings Algorithm}\label{sec:algo} 

The Metropolis-Hastings algorithm makes use of a proposal density or random generator $q(y|x)$ which might be a simple distribution such as normal. For this study purposes, the term $q(y|x)$ will be applied to a set of three mutually exclusive random generators presented in Section \ref{sec:RA}: $q(y|x)=$(Brownian, Illusion, Reservoir). 

The key idea of this alternative version of the MHRW algorithm is to generate a number of independent samples from a given random generator. Thus, it is necessary to sample a candidate node $\varrho$ from $q(y|x)=$(Brownian, Illusion, Reservoir). The candidate node is accepted if and only if this node belongs to the graph $G$. Technically speaking, $G$ needs to be stored as a data structure enabled to handle the graph evolution in time. The steps of this method are outlined in algorithm \ref{alg:metroT}.

\begin{algorithm}
 \DontPrintSemicolon
 \SetAlgoLined
 initialization:\; 
 $t \leftarrow 0$ initial time;\;  
 $v_0 \leftarrow 0$ initial node;\;   
 $q(y|x) \leftarrow $ Brownian or Illusion or Reservoir;\;   
 $\varrho \leftarrow q(y|x)$ a candidate node from a given random generator;\;\; 
 
 \While{stopping criterion not met}{
\;
Generate a candidate node $\varrho$ from $q(y|x)$; 

\;

\eIf{$\varrho \in G$}{
\;
$v \leftarrow \varrho$;
}
{
\;
Stay at $v$;
}
Set $t = t+1$; 
}
 
\caption{Alternative version of the Metropolis-Hastings Random Walk Algorithm.}
\label{alg:metroT}
\end{algorithm}

It should be highlighted that the node $v_0$ is placed in terms of the first record retrieved from the servers of Twitter. Similarly, the stopping criterion is determined by the number of countries randomly chosen e.g., it is possible to select 15 countries with their respective top 10 trending topics (15 $\times$ 10 trends in total). The next section will describe how these countries are obtained.

\section{Sampling Global Trends on Twitter}\label{sec:sampling}
\subsection{Pre-processing}

In this stage, relevant queries are sent to the social media service via the Twitter API to extract and monitor social information \cite{abdesslem2012reliable}. Therefore, it has been developed a content extraction tool on Java called ``social explorer'', which intends to discover global trends on Twitter. Thus, the social explorer is integrated with the Twitter service through a Java library called: \texttt{Twitter4J}, which is a simple and flexible Java library for the Twitter API. This open source software can be downloaded from: \url{http://twitter4j.org/}.

The social explorer is able to connect itself to the streams of the public data flowing through Twitter. In terms of the Twitter API connection, this application makes a request, then a server from Twitter opens and accepts a streaming connection and finally this server pulls processed results from a data store.

For the estimation of trends concentration, a list of countries with publicly available trends was requested from Twitter. Countries are identified by means of a specific \texttt{WOEID}. The term \texttt{WOEID} refers to a service that allows to look up the unique identifier called the ``Where on Earth ID'' (see \url{http://developer.yahoo.com/geo/geoplanet/}).

To establish which countries had more activity on Twitter, a range of WOEIDs was chosen between $23424000$ and $23425000$. This range was determined according to a set of empirical trials. A full list of retrieved countries can be found in Table \ref{tab:Countnames}.

\begin{table}[!htb]
\centering
\caption{Table of retrieved countries with more activity on Twitter during our empirical trials.}
\begin{tabular}{|l|l| }
\hline
\multicolumn{2}{|c|}{\textbf{List of Countries}} \\
\hline
\hline
Argentina&Australia\\
Belgium&Brazil\\
Canada	&Chile	\\
Colombia &Dom. Republic\\
Ecuador	& France\\
Germany	&Greece	\\
Guatemala	&India\\
Indonesia	&Ireland\\
Italy	&Japan	\\
Kenya	&Korea	\\
Malaysia &Mexico\\
Netherlands	&New Zealand	\\
Nigeria	&Norway	\\
Pakistan	&Peru	\\
Philippines	&Poland	\\
Portugal	&Russia	\\
Singapore	&South Africa \\
Spain	&Sweden	\\
U. Arab Emirates & Turkey	\\
Ukraine	& United Kingdom\\
United States	&Venezuela	\\
\hline
\end{tabular}
\label{tab:Countnames}
\end{table}

The algorithm \ref{alg:metroT} interacts with Twitter via its public API as a primary way to retrieve data. Once all the information has been retrieved, a random sampling is performed across the global trends using Algorithm \ref{alg:metroT}. Collected samples are stored in an output data file and depicted on a visual interface.

A sample is chosen according to the following eligibility criteria (initial conditions): 1) number of countries and 2) a minimum number of users following a global trend. In this study, the initial conditions consisted of 15 countries and 10 users. Therefore, a maximum of 150 (15 $\times$ 10) trending topics per independent run were available to be gathered. However, due to a trending topic or an user can be counted multiple times, which makes the measurement hard to interpret, all duplicate trends and duplicate users were removed from the sample. After filtering out all duplicates, it can be built a data structure containing a set of unique records.

In summary, the steps to generate the data are as follows:

\begin{enumerate}
	\item Collect a list of \texttt{WOEIDs} by searching countries with publicly available trends, then select randomly a set of $W=15$ unique countries. 
	\item for each country $c \in W$, we acquire a list of the top ten trending topics($TT$), add each trending topic $TT$ as a node to the graph $G$. Then, set the minimum number  of users following this trend to $Fr=10$;
	\item for each $TT$, get a list of users linked to the corresponding trending topic, e.g. $Fr(TT)$, and add each of them as a node to $G$;
	\item create an edge $[TT,Fr(TT)]$ and add it to $G$; 	
	\item save the graph $G$. 	
\end{enumerate}

\section{Results}\label{sec:results}

In order to assess the performance of the social explorer, a sample of publicly available trends was collected, this random sample contains tweets posted from December 17 to December 20 2013, between 16:30 and 22:30 GMT (time window). This sample consisted of $3,325$ trending topics generated by $225,102$ unique users that emerged during the observed time window. 

It is important to note that in this case, not only tweets written in English were extracted. This feature provides a different framework with respect to previous studies whereby only English tweets were collected e.g., \cite{Lilian2013srep,weng2012competition}. One advantage of this multilingual feature is that it avoids a bias in terms of the information posted in English. 

To replicate the sampling process, a series of $10$ independent walks was performed for each one of the three  random generators: $q(y|x)=$(Brownian, Illusion, Reservoir) ($30$ runs in total). Then, two different output files were stored for further analysis: a \texttt{.dat} file and a \texttt{.gml} file. The first one contains information such as: Total number of trending topics, total number of unique followers, number of iterations, total number of sampled trends, a full list of the collected trends, number of nodes, number of edges, node degree per trending topic, memory usage, total number of duplicates and the elapsed time during the sampling process. On the other hand, the second output file contains a GML (Graph Modeling Language) formatted file, which describes a graph obtained by the social explorer. This file is used to build and evaluate graphically each one of the samples.   

\begin{figure}[!htb]
\centering
\includegraphics[scale=.55]{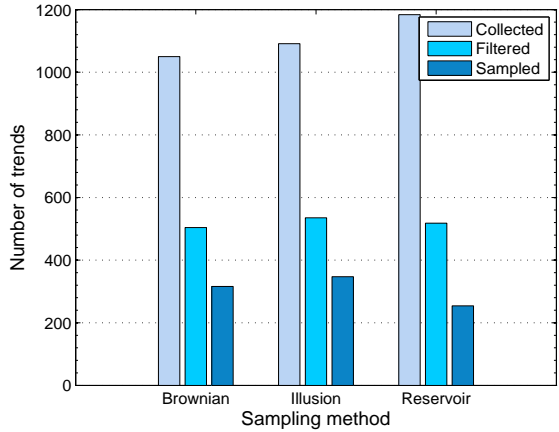}
\caption{Plot divided into three main criteria: number of trends retrieved from Twitter (collected);  number of trends after removing all duplicates (filtered) and  number of sampled trends collected by each random generator (sampled).}
\label{fig:AllDat}
\end{figure}

\begin{figure}[!htb]
\centering
\includegraphics[scale=.55]{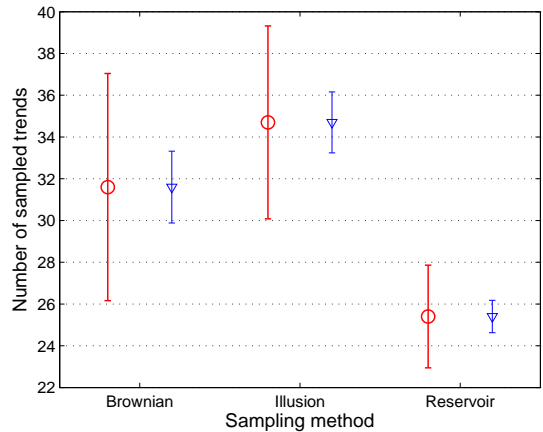}
\caption{Means  corresponding to the average of sampled trends.}
\label{fig:SEtre}
\end{figure}

Fig.\ref{fig:AllDat} compares a cumulative analysis of the number of trends. This plot may be divided into three main criteria: number of trends retrieved from the Twitter service (collected), number of trends after removing all duplicates (filtered) and number of sampled trends collected by each random generator (sampled). Similarly, means with respect to the number of sampled trends are shown in Fig. \ref{fig:SEtre}. What is interesting in this data, is that the sampled trends represents the core information to evaluate how was the three models behavior in terms of data collection. It should be highlighted that the number of collected trends depends exclusively from the Twitter service. Likewise, the filtering process was carried out as a cleaning data process. 

\begin{figure}[!htb]
\centering
\includegraphics[scale=.55]{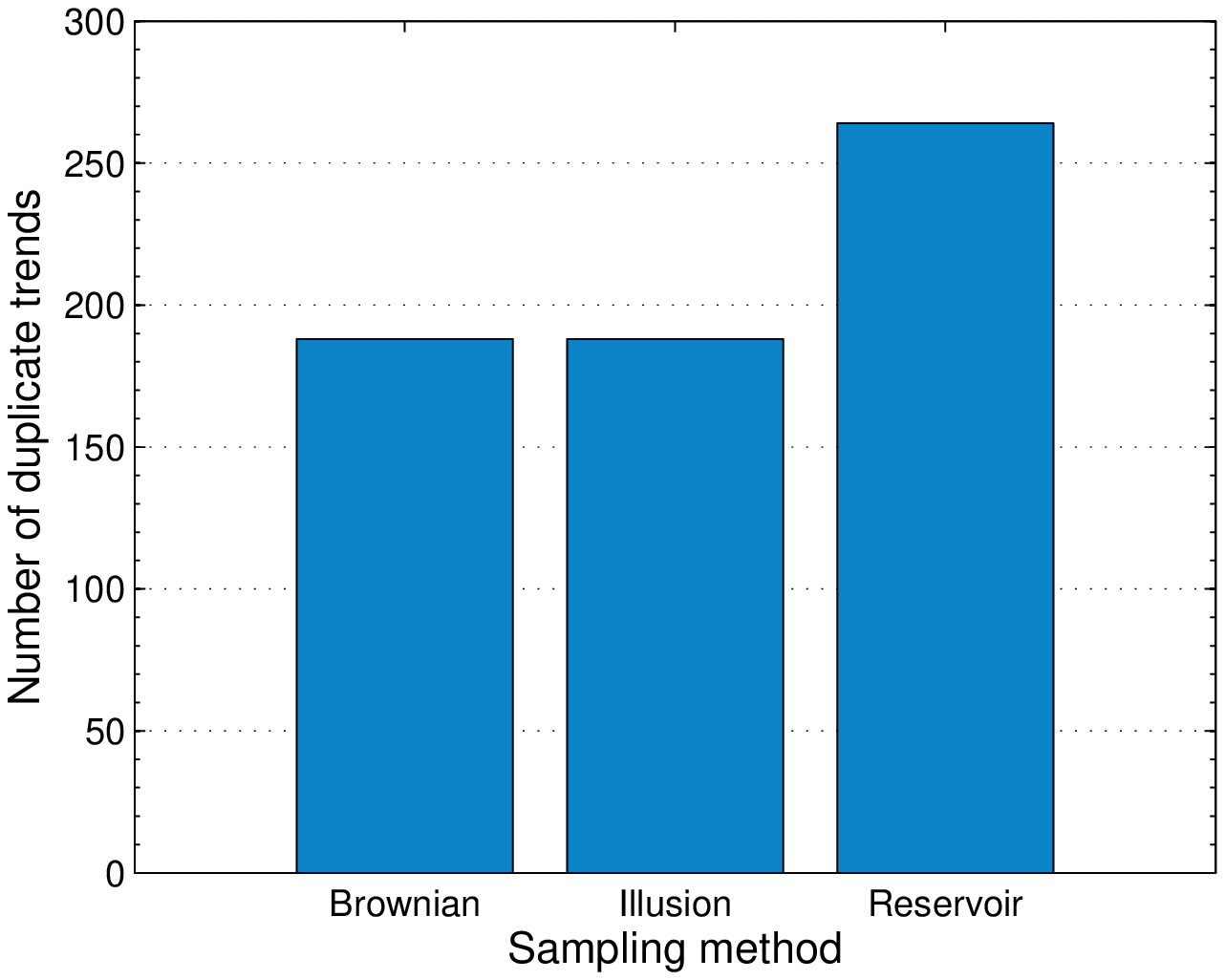}
\caption{Number of duplicate trends for $q(y|x)$.}
\label{fig:Bardup}
\end{figure}

\begin{figure}[!htb]
\centering
\includegraphics[scale=.55]{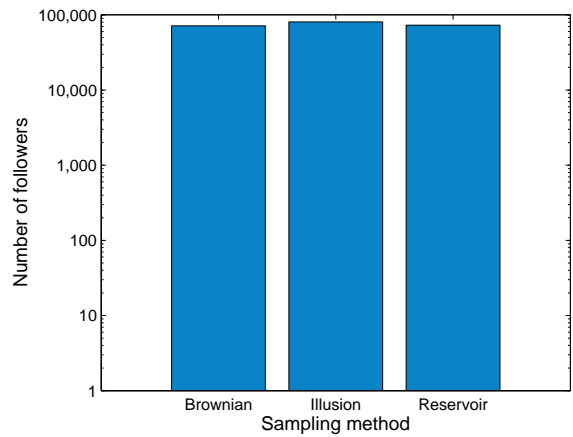}
\caption{Number of unique followers presented with logarithmic scale for the y-axis.}
\label{fig:BarFo}
\end{figure}

Owing to the natural tendency of the social explorer to move toward the same node many times, which is induced by each one of the random generators, a considerable number of duplicate trends is added to the output sequence. This permits to compare the results in terms of the number of duplicate trends generated during the observation time window (see Fig. \ref{fig:Bardup}). Likewise, Fig. \ref{fig:BarFo} compares the number of unique followers obtained from each random generator $q(y|x)$. 

\begin{figure}[!htb]
\centering
\includegraphics[scale=.55]{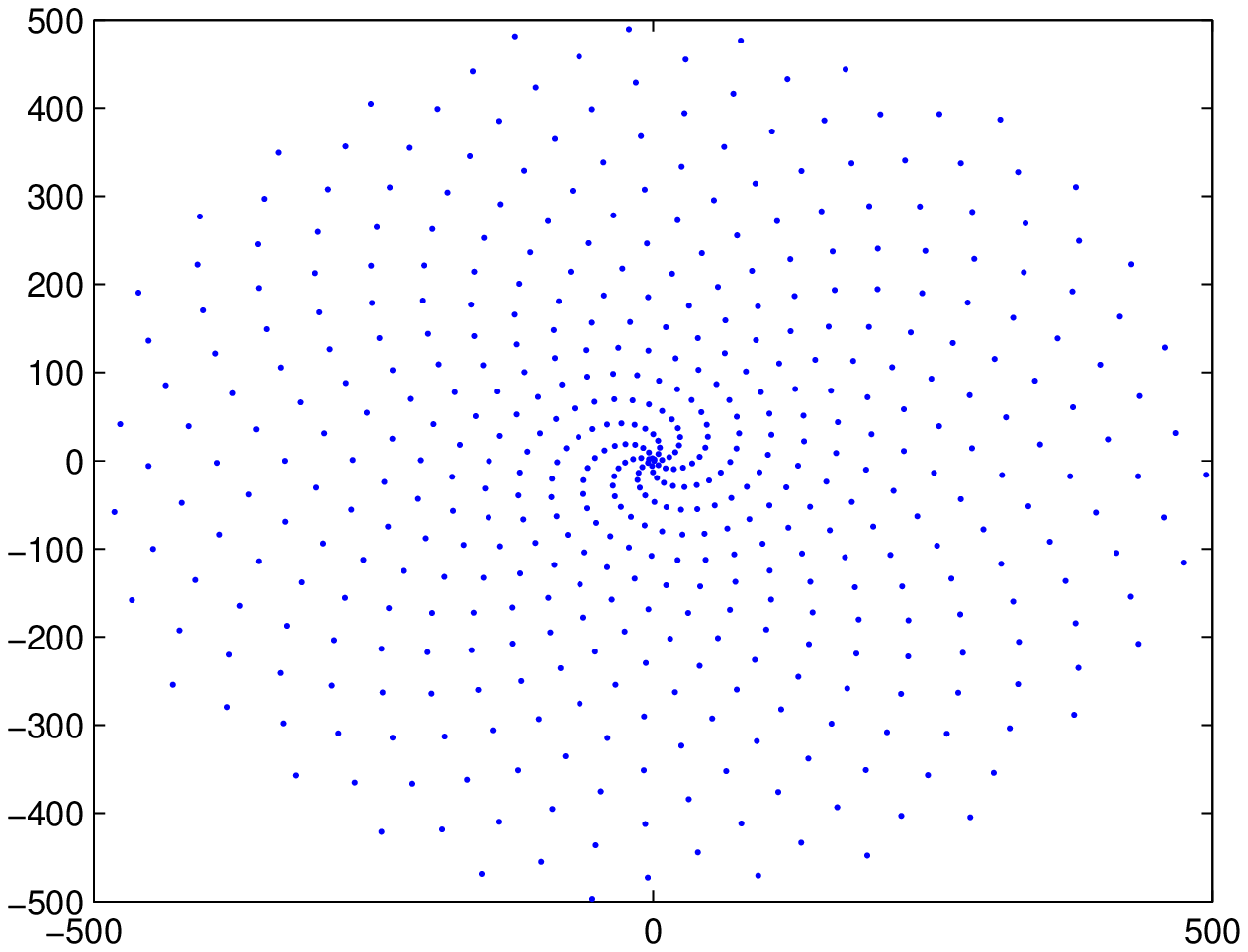}
\caption{Pattern visualization of the \textit{Illusion} spiral.}
\label{fig:illu}
\end{figure}

Table \ref{tab:DS} compares a descriptive statistics of the average number in terms of: trends filtered, sampled trends, duplicate trends and the number of followers. It is apparent from this table that the Illusion model slightly differs from the rest of the random generators in terms of sampled trends. This difference may be caused by the spread-out pattern presented on the shape of this spiral (see Fig. \ref{fig:illu}). 

\begin{table}[!htb]
\centering
\caption{Descriptive Statistics during the observation time window.}
\begin{tabular}{ |l|l|l|l|l| }
\hline
\hline
Trends & $q(y|x)$ & total & avg & std  \\ 
\hline
\hline
\multirow{3}{*}{Filtered} 
 & Brownian &504 &50.4 &8.16 \\
 & Illusion &535 &53.5 &7.59 \\
 & Reservoir &518 &51.8 &4.07  \\  
 \hline 
 \hline
\multirow{3}{*}{Sampled} 
 & Brownian &316 &31.60 &5.44 \\
 & Illusion &347 &34.70 &4.62 \\
 & Reservoir &254 &25.40 &2.45  \\ 
 \hline 
 \hline
\multirow{3}{*}{Duplicated}
 & Brownian &188 &18.80 &3.08 \\
 & Illusion &188 &18.80 &3.04 \\
 & Reservoir &264 &26.40 &2.91  \\ 
\hline 
\hline
\multirow{3}{*}{Followers}
 & Brownian &71574 &7157.4 &1696.6 \\
 & Illusion &80522 &8052.2 &2654.8 \\
 & Reservoir &73006 &7300.6 &1582  \\ 
\hline
\end{tabular}
\label{tab:DS}
\end{table}

\begin{figure}[!htb]
\centering
\subcaptionbox{Brownian.} 
{\includegraphics[scale=.55]{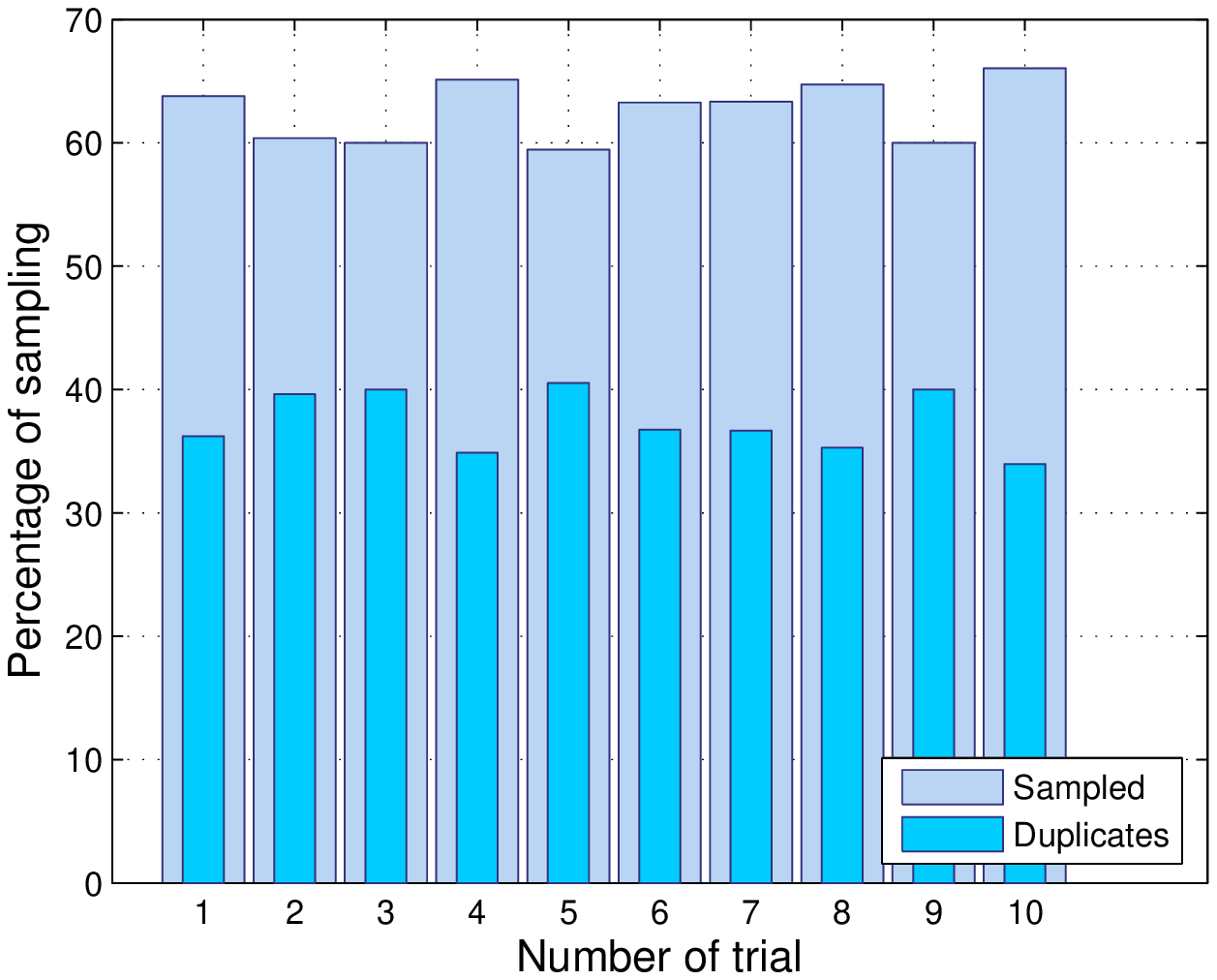}}
\subcaptionbox{Illusion.}
{\includegraphics[scale=.55]{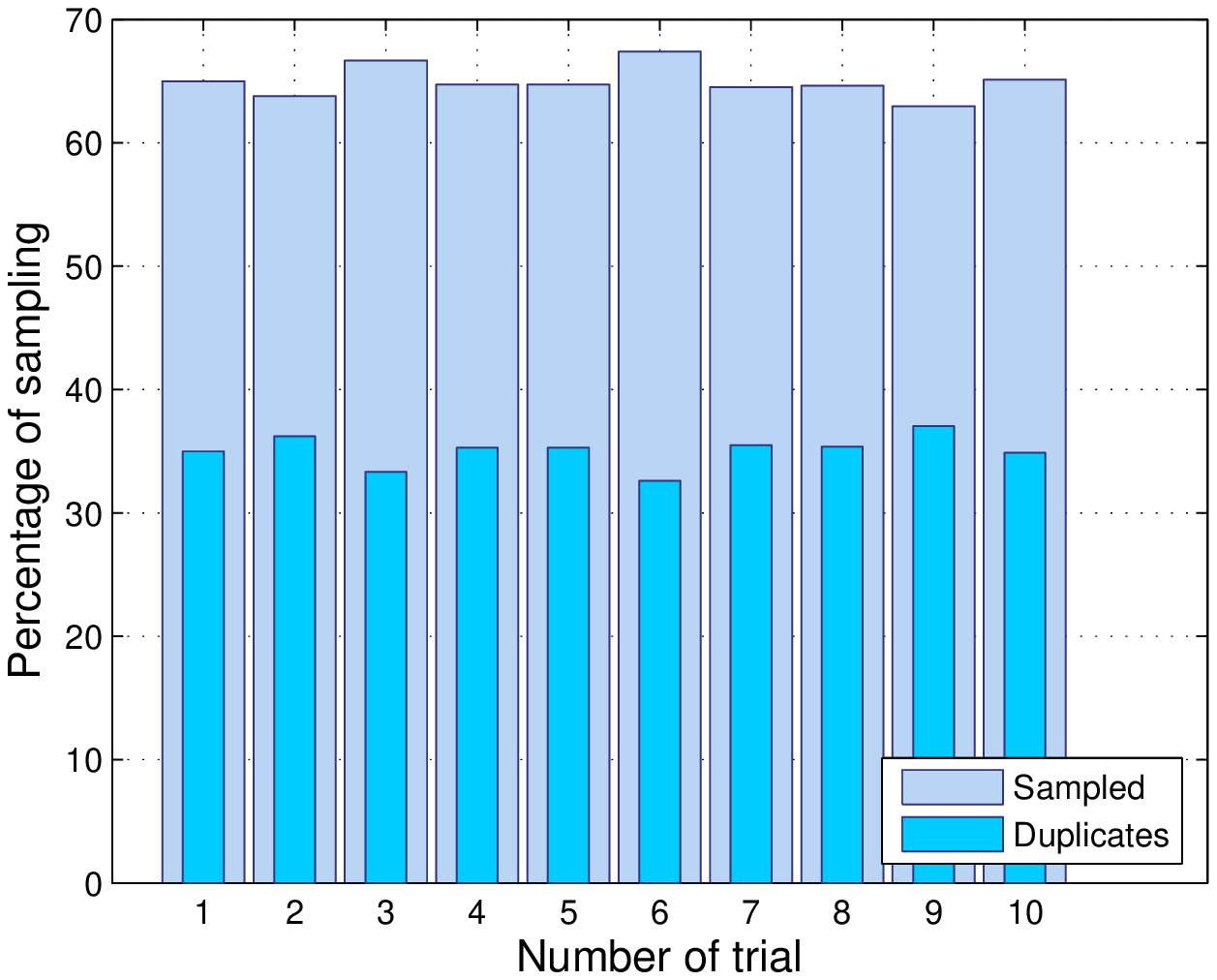}}
\subcaptionbox{Reservoir.}
{\includegraphics[scale=.55]{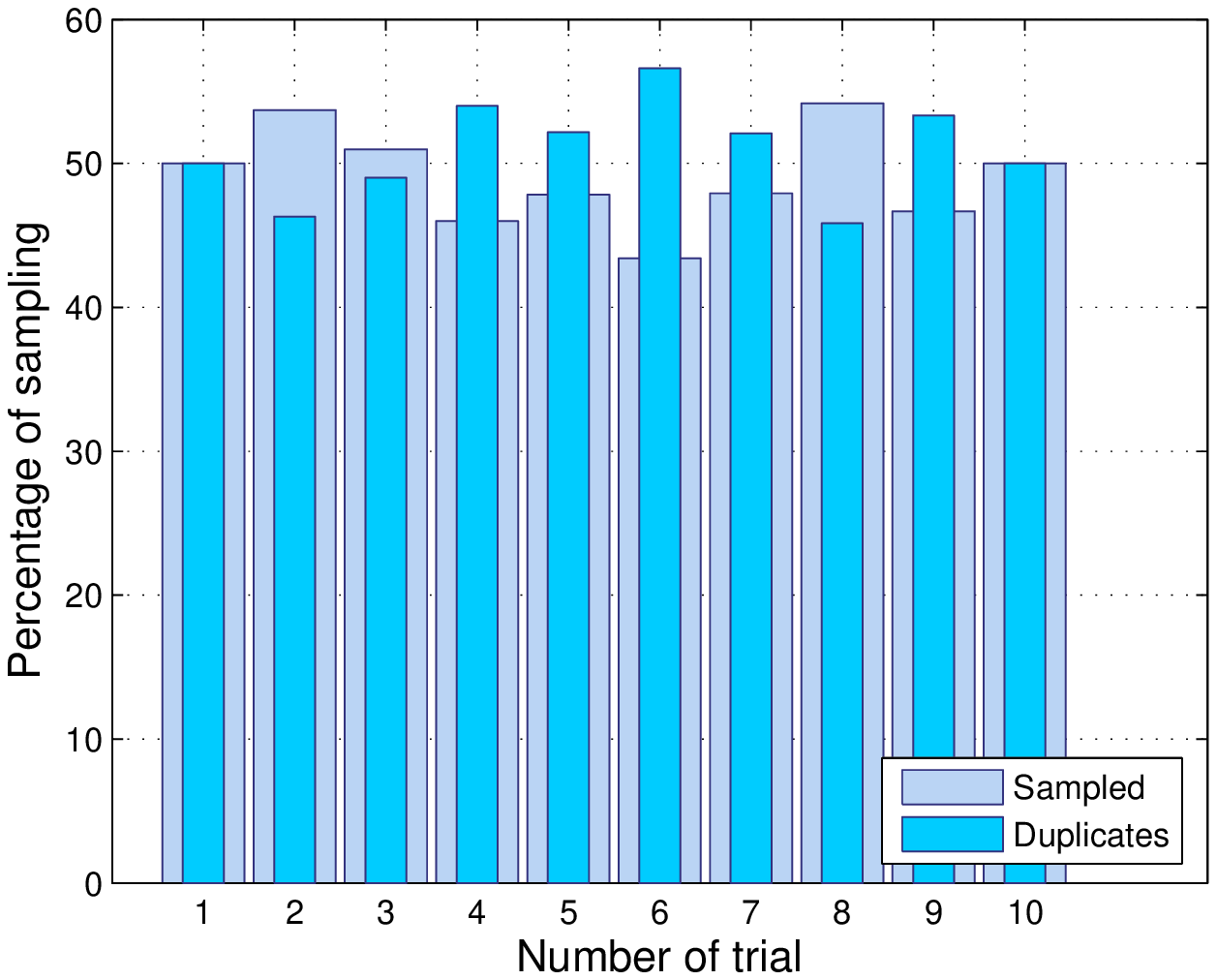}}
\caption{Group of plots of the percentage of accuracy plotted versus the number of trials. The measures are computed based on the percentage of sampled trends and on the percentage of duplicate trends generated by each random generator.    
\label{fig:percen}}
\end{figure}

In addition, the percentage of accuracy was computed based on the ratio obtained by dividing the cumulated value of sampled trends by the total number of iterations carried out by the social explorer. The same procedure was applied to the cumulated value of duplicate trends. The results obtained from this analysis are summarized in Fig.\ref{fig:percen}. This set of plots considers the percentage amounts of sampled and duplicate trends over all the 30 samples produced by each random generator. Data from this figure can be compared with the data in Table \ref{tab:perc} which shows that the Brownian and Illusion models perform well during their 10 independent runs. However, the reservoir model showed a poor performance in terms of percentage of sampling.

Basic statistics of the average amount of the relative accuracy in sampling and the average amount of the percentage of duplicate trends, are reported in Table \ref{tab:perc}. 

\begin{table}[!htb]
\centering
\caption{Descriptive Statistics of the average amount of sampled and duplicate trends. It can be seen from the data that the Illusion spiral outperforms all others random generators in terms of the number of sampled trends.}
\begin{tabular}{ |l|l|l|}
\hline
\hline
Trends & $q(y|x)$ & avg  \\ 
\hline
\hline
\multirow{3}{*}{ \% Sampled} 
 & Brownian & 62.61\% \\
 & Illusion & 64.94\% \\
 & Reservoir &49.06\%   \\  
 \hline 
 \hline
\multirow{3}{*}{ \% Duplicated} 
 & Brownian & 37.3\%\\
 & Illusion & 35.05\% \\
 & Reservoir & 50.93\% \\ 
\hline 
\end{tabular}
\label{tab:perc}
\end{table}

\subsection{Memory consumption}

This section examines the estimated memory usage employed by each proposed model. The results obtained from the preliminary analysis of memory consumption can be compared in Table \ref{tab:Mem}, this table compares the average memory consumption in Megabytes (MB) and the total of memory used across 10 independent runs. In this regard, there were no significant differences between the amount of MB used for each random generator. 

\begin{table}[!htb]
\centering
\caption{Descriptive Statistics of memory consumption. This table compares the average memory consumption in Megabytes (MB) and the total of memory used across 10 independent runs.}
\begin{tabular}{|l|l|l|l|l|}
\hline
\hline
Results & $q(y|x)$ & total & avg & std  \\ 
\hline
\hline
\multirow{3}{*}{Memory (MB)} 
 & Brownian &66 &6.6 &4.16 \\
 & Illusion &66 &6.6 &2.71 \\
 & Reservoir &62 &6.2 &4.61   \\  
\hline 
\end{tabular}
\label{tab:Mem}
\end{table}

\begin{figure}[!htb]
\centering
\includegraphics[scale=.55]{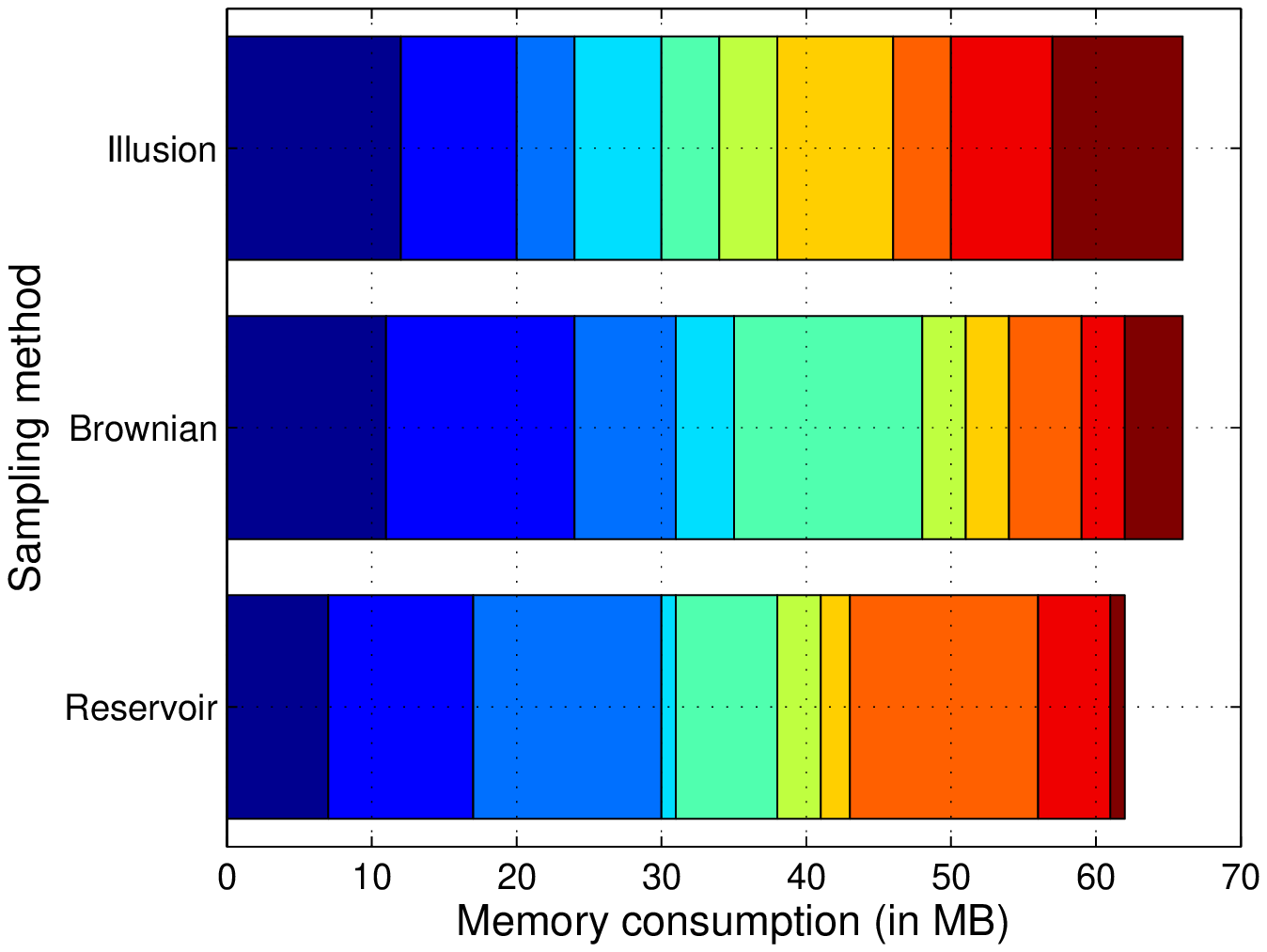}
\caption{Stacked bar chart displaying the sum of the memory consumption split in 10 independent runs per random generator.
\label{fig:mem}}
\end{figure}

Due to the experiments were run using custom software written in Java, it has been considered to assess the results obtained related to the memory consumption in Megabytes (MB). The basic computer hardware information is as follows:  Processor: Intel(R) Core(TM)2 Duo CPU at 3.33GHz. Installed memory (RAM): 4.00 GB. System type: 64-bit operation system. The application was run on Windows 7 Enterprise edition. Fig \ref{fig:mem} presents a cumulative memory usage plot. This plot is presented as a stacked bar which provides the sum of all the memory consumption across 10 independent walks per model. From these data, it can be seen that there were no significant differences between the sampling methods used as random generators. 

\subsection{Convergence Monitoring}

Part of the aim of this research is to identify convergence during the sampling process. Therefore, a convergence analysis was prepared according to the procedure used by the Geweke to evaluate the accuracy of sampling-based approaches \cite{geweke1991evaluating,lee2006statistical}. This Geweke diagnostics is a standard Z-score which consists in taking two non-overlapping parts of the Markov chain and compares the means of both parts, using a difference of means test to see if the two parts of the chain are from the same distribution (null hypothesis).

This diagnostic represents a test of whether the sample of draws has attained an equilibrium state based on the first $10\%$ of the sample of draws, versus the last $50\%$ of the sample of draws. If the Markov chain of draws has reached an equilibrium state, it would be expected to obtain roughly equal averages from these two splits of the sample \cite{lesage1999applied}. MATLAB functions can be found at \url{http://www.spatial-econometrics.com/gibbs/contents.html}.

\begin{figure}[!htb]
\centering
\subcaptionbox{Brownian.} 
{\includegraphics[scale=.60]{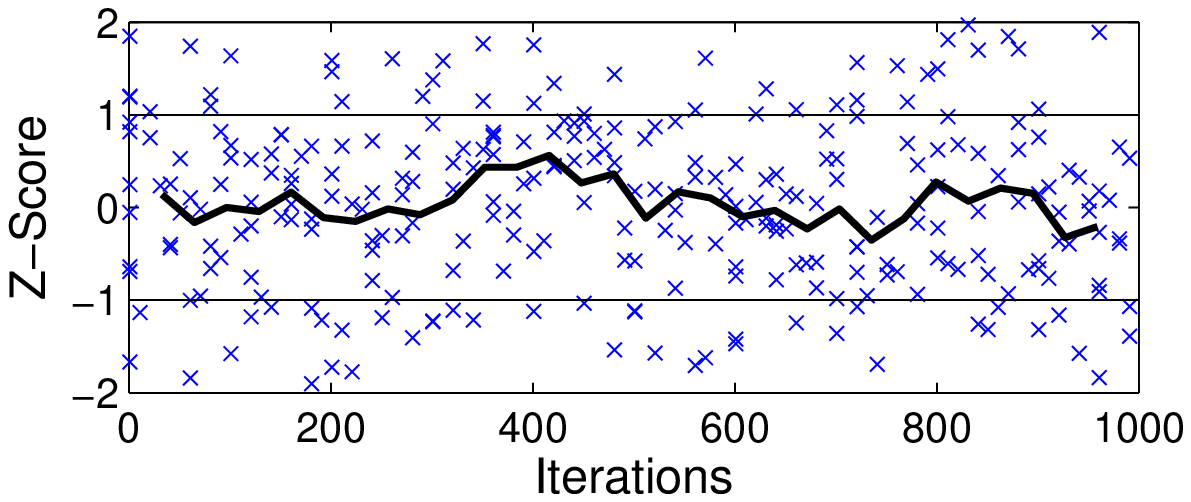}}
\subcaptionbox{Illusion.}
{\includegraphics[scale=.60]{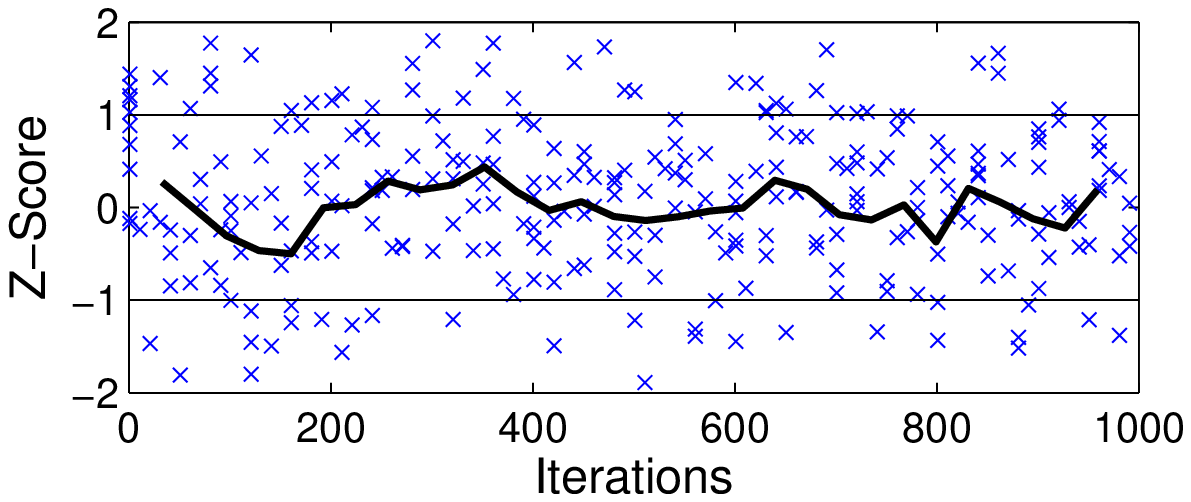}}
\subcaptionbox{Reservoir.}
{\includegraphics[scale=.60]{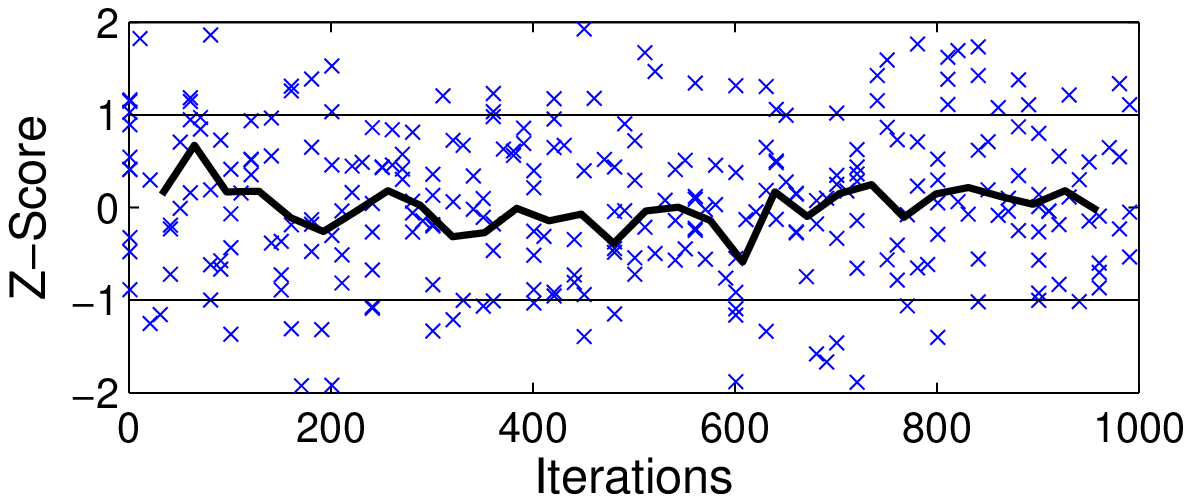}}
\caption{Plots of the resulting Z-scores against the number of iterations for the metric of node degree (number of users that follow a particular trend). Horizontal lines at $Z=\pm1$  are added to the plots to indicate the convergence interval.
\label{fig:geweke2}}
\end{figure}

Fig.\ref{fig:geweke2} provides trace plots for the property of node degree (number of users that follow a particular trend). These plots present the Z-score value against the number of iterations. Therefore, using the Geweke diagnostics it is possible to identify the convergence analysis for the Brownian walk, the Illusion spiral and the Reservoir sampling. The number of draws was fixed to $1100$ with a \textit{ burn-in} process discarding the first $100$. Thus, in accordance to \cite{gjoka2011practical} we can  declare convergence when most values fall in the [-1, 1] interval. Additionally, we plot an average line using 30 points on the x-axis.  Finally, as it can be seen in Fig. \ref{fig:geweke2} our convergence analysis suggests that our sample draws have attained an equilibrium state showing that the means of the values converge rapidly in the sequence.

\section{Limitations}\label{sec:limit}

As we mentioned before in Section \ref{sec:results}, one advantage of this approach is the multilingual feature which avoids a bias in terms of the information posted in English. However, there are certain drawbacks associated with the use of different languages e.g., lack of knowledge of the language and the misinterpretation of the statements. 

On the other hand, this research does not take into account that the social explorer is not able to distinguish between Twitterbots\footnote{A Twitterbot is a program used to produce automated posts on the Twitter microblogging service, or to automatically follow Twitter users.} and real users on Twitter. Therefore, all the estimates include Twitterbots causing an over estimation in the results. These data must be interpreted with caution since all the information collected from this study is mainly based on the Twitter response service i.e., it is possible to find some inconsistencies in terms of the outcomes retrieved through the custom software written in Java.

\section{Conclusions}\label{sec:con}

This paper has explained the central importance of defining a standard sampling methodology applicable to cases where the social network information flow is readily available. The main purpose of the current study was to assess a low computational cost method for sampling emerging global trends on Twitter.  

It is now possible to state that the development of a faster randomized algorithm able to carry out a collecting process via three random generators, can be effectively gathered by using $q(y|x)=\{Brownian,Illusion,Reservoir\}$. The present paper confirms previous findings related to the good performance of the Brownian and Illusion generators \cite{pina2013collecting}. It should be noted that according to the first systematic study of using a Metropolis-Hastings Random Walk (MHRW) reported by Gjoka \textit{et al.} in \cite{gjoka10_walkingfb}, where the authors demonstrated that the MHRW perform well under Facebook using a normal distribution as a random generator (a Brownian walk in this study), the results of this research show that the MHRW, is eligible to be modified in terms of how it heuristically generates a candidate node using different random generators such as:  Illusion and Reservoir.

The empirical findings of this study suggest that, sampling global trends on Twitter has several practical applications related to extract real-time information. Despite its exploratory nature by looking at how impactful people are about a specific topic and within specific categories, this research offers some insight into how to collect publicly available trends using a social explorer, which works as an interface between a faster randomized algorithm proposed in Algorithm \ref{alg:metroT} and Twitter. 

Overall, our current study indicates that our sampling methodology may be a promising new approach to social networking service analysis and an useful exploration tool for social data acquisition.

\section{Future Work}\label{sec:future}

As online social networks can provide a good idea about how content spreads, future work should therefore concentrate on the investigation of social media monitoring, to map leading indicators and see why some events or trends might be triggered. In this regard, it would be interesting to assess the effects of constantly scan different public web sources including news publications. On the other hand, it is recommended that further research be undertaken in the following areas: cyber threat intelligence, sentiment analysis, smart content marketing and political topics. These areas could produce an interesting social intelligence platform able to detect and predict emerging threats. In addition, it could be helpful to present graphic reports showing these events or trends sorted chronologically or by importance. 

A sentiment analysis approach would help to establish a greater degree of accuracy to determine the attitude of a social network user, with respect to some trending topic. Thus, these findings suggest several courses of action for future practice. Another natural progression of this work is to analyze the impact of different social networks on countries where Twitter has been banned e.g., China. In this regard, it becomes very important to consider topics in different languages such as: mandarin or Russian. Two major online social networks have appeared on the map to fill the gap in the Asian continent. These are: Sina Weibo \url{http://www.weibo.com/} and VK \url{http://vk.com/}. Both social networks provide an open API to access to the respective information. Thus, further investigation and experimentation into these two online social networks is strongly recommended.

\bibliographystyle{plain}
\bibliography{RefBib}

\begin{thebibliography}{10}

\bibitem{abdesslem2012reliable}
Fehmi~Ben Abdesslem, Iain Parris, and Tristan Henderson.
\newblock Reliable online social network data collection.
\newblock In {\em Computational Social Networks}, pages 183--210. Springer,
  2012.

\bibitem{backstrom2011supervised}
Lars Backstrom and Jure Leskovec.
\newblock Supervised random walks: predicting and recommending links in social
  networks.
\newblock In {\em Proceedings of the fourth ACM international conference on Web
  search and data mining}, pages 635--644. ACM, 2011.

\bibitem{bar2008random}
Ziv Bar-Yossef and Maxim Gurevich.
\newblock Random sampling from a search engine's index.
\newblock {\em Journal of the ACM (JACM)}, 55(5):24, 2008.

\bibitem{berg1993random}
H.C. Berg.
\newblock {\em {Random walks in biology}}.
\newblock Princeton Univ Pr, 1993.

\bibitem{bhattacharyya2011analysis}
Prantik Bhattacharyya, Ankush Garg, and Shyhtsun~Felix Wu.
\newblock Analysis of user keyword similarity in online social networks.
\newblock {\em Social network analysis and mining}, 1(3):143--158, 2011.

\bibitem{caci2011facebook}
Barbara Caci, Maurizio Cardaci, and Marco~E Tabacchi.
\newblock Facebook as a small world: a topological hypothesis.
\newblock {\em Social Network Analysis and Mining}, pages 1--5, 2011.

\bibitem{caci2012facebook}
Barbara Caci, Maurizio Cardaci, and Marco~E Tabacchi.
\newblock Facebook as a small world: a topological hypothesis.
\newblock {\em Social Network Analysis and Mining}, 2(2):163--167, 2012.

\bibitem{cipra2000best}
Barry~A Cipra.
\newblock The best of the 20th century: Editors name top 10 algorithms.
\newblock {\em SIAM news}, 33(4):1--2, 2000.

\bibitem{codling2010diffusion}
E.A. Codling, R.N. Bearon, and G.J. Thorn.
\newblock {Diffusion about the mean drift location in a biased random walk}.
\newblock {\em Ecology}, 91(10):3106--3113, 2010.

\bibitem{davis1993spirals}
P.J. Davis, W.~Gautschi, and A.~Iserles.
\newblock {\em Spirals: from Theodorus to chaos}.
\newblock AK Peters, 1993.

\bibitem{dudewicz1976introduction}
E.J. Dudewicz.
\newblock {\em Introduction to statistics and probability}.
\newblock Holt, Rinehart and Winston, 1976.

\bibitem{ferri2012new}
Fernando Ferri, Patrizia Grifoni, and Tiziana Guzzo.
\newblock New forms of social and professional digital relationships: the case
  of facebook.
\newblock {\em Social Network Analysis and Mining}, 2(2):121--137, 2012.

\bibitem{fire2013organization}
Michael Fire, Rami Puzis, and Yuval Elovici.
\newblock Organization mining using online social networks.
\newblock {\em arXiv preprint arXiv:1303.3741}, 2013.

\bibitem{geweke1991evaluating}
J.~Geweke et~al.
\newblock {\em Evaluating the accuracy of sampling-based approaches to the
  calculation of posterior moments}.
\newblock Federal Reserve Bank of Minneapolis, Research Department, 1991.

\bibitem{gjoka2011practical}
M.~Gjoka, M.~Kurant, C.T. Butts, and A.~Markopoulou.
\newblock Practical recommendations on crawling online social networks.
\newblock {\em Selected Areas in Communications, IEEE Journal on},
  29(9):1872--1892, 2011.

\bibitem{gjoka2011multigraph}
Minas Gjoka, Carter~T Butts, Maciej Kurant, and Athina Markopoulou.
\newblock Multigraph sampling of online social networks.
\newblock {\em Selected Areas in Communications, IEEE Journal on},
  29(9):1893--1905, 2011.

\bibitem{gjoka10_walkingfb}
Minas Gjoka, Maciej Kurant, Carter~T. Butts, and Athina Markopoulou.
\newblock {Walking in Facebook: A Case Study of Unbiased Sampling of OSNs}.
\newblock In {\em Proceedings of IEEE INFOCOM '10}, San Diego, CA, March 2010.

\bibitem{golbeck2013analyzing}
Jennifer Golbeck.
\newblock {\em Analyzing the social web}.
\newblock Newnes, 2013.

\bibitem{hawelka2013geo}
Bartosz Hawelka, Izabela Sitko, Euro Beinat, Stanislav Sobolevsky, Pavlos
  Kazakopoulos, and Carlo Ratti.
\newblock Geo-located twitter as the proxy for global mobility patterns.
\newblock {\em arXiv preprint arXiv:1311.0680}, 2013.

\bibitem{kallus2014predicting}
Nathan Kallus.
\newblock Predicting crowd behavior with big public data.
\newblock {\em arXiv preprint arXiv:1402.2308}, 2014.

\bibitem{kurant11_magnifying}
Maciej Kurant, Minas Gjoka, Carter~T. Butts, and Athina Markopoulou.
\newblock {Walking on a Graph with a Magnifying Glass: Stratified Sampling via
  Weighted Random Walks}.
\newblock In {\em Proceedings of ACM SIGMETRICS '11}, San Jose, CA, June 2011.

\bibitem{lee2006statistical}
Sang~Hoon Lee, Pan-Jun Kim, and Hawoong Jeong.
\newblock Statistical properties of sampled networks.
\newblock {\em Physical Review E}, 73(1):016102, 2006.

\bibitem{lesage1999applied}
J.P. LeSage.
\newblock Applied econometrics using matlab.
\newblock {\em Manuscript, Dept. of Economics, University of Toronto}, 1999.

\bibitem{lin2013bigbirds}
Yu-Ru Lin, Drew Margolin, Brian Keegan, Andrea Baronchelli, and David Lazer.
\newblock \# bigbirds never die: Understanding social dynamics of emergent
  hashtag.
\newblock {\em arXiv preprint arXiv:1303.7144}, 2013.

\bibitem{lu2014network}
Xin Lu and Christa Brelsford.
\newblock Network structure and community evolution on twitter: Human behavior
  change in response to the 2011 japanese earthquake and tsunami.
\newblock {\em Scientific reports}, 4, 2014.

\bibitem{maiya2010sampling}
Arun~S Maiya and Tanya~Y Berger-Wolf.
\newblock Sampling community structure.
\newblock In {\em Proceedings of the 19th international conference on World
  wide web}, pages 701--710. ACM, 2010.

\bibitem{martinez2001computational}
W.L. Martinez and A.R. Martinez.
\newblock {\em Computational statistics handbook with MATLAB}, volume~2.
\newblock Chapman \& Hall/CRC, 2001.

\bibitem{mislove2006exploiting}
Alan Mislove, Krishna~P Gummadi, and Peter Druschel.
\newblock Exploiting social networks for internet search.
\newblock In {\em 5th Workshop on Hot Topics in Networks (HotNets06).
  Citeseer}, page~79, 2006.

\bibitem{mitchell2013geography}
Lewis Mitchell, Morgan~R Frank, Kameron~Decker Harris, Peter~Sheridan Dodds,
  and Christopher~M Danforth.
\newblock The geography of happiness: Connecting twitter sentiment and
  expression, demographics, and objective characteristics of place.
\newblock {\em PloS one}, 8(5):e64417, 2013.

\bibitem{pina2013collecting}
CA~Pina-Garcia and Dongbing Gu.
\newblock Collecting random samples from facebook: An efficient heuristic for
  sampling large and undirected graphs via a metropolis-hastings random walk.
\newblock In {\em Systems, Man, and Cybernetics (SMC), 2013 IEEE International
  Conference on}, pages 2244--2249. IEEE, 2013.

\bibitem{pina2013spiraling}
CA~Pi{\~n}a-Garc{\'\i}a and Dongbing Gu.
\newblock Spiraling facebook: an alternative metropolis--hastings random walk
  using a spiral proposal distribution.
\newblock {\em Social Network Analysis and Mining}, 3(4):1403--1415, 2013.

\bibitem{robert2009introducing}
C.~Robert and G.~Casella.
\newblock {\em Introducing Monte Carlo Methods with R}.
\newblock Springer, 2009.

\bibitem{scott2011social}
John Scott.
\newblock Social network analysis: developments, advances, and prospects.
\newblock {\em Social network analysis and mining}, 1(1):21--26, 2011.

\bibitem{takhteyev2012geography}
Yuri Takhteyev, Anatoliy Gruzd, and Barry Wellman.
\newblock Geography of twitter networks.
\newblock {\em Social Networks}, 34(1):73--81, 2012.

\bibitem{thapen2013towards}
Nicholas~A Thapen and Moustafa~M Ghanem.
\newblock Towards passive political opinion polling using twitter.
\newblock In {\em BCS SGAI SMA 2013 The BCS SGAI Workshop on Social Media
  Analysis}, page~19, 2013.

\bibitem{ugander2011anatomy}
J.~Ugander, B.~Karrer, L.~Backstrom, and C.~Marlow.
\newblock The anatomy of the facebook social graph.
\newblock {\em Arxiv preprint arXiv:1111.4503}, 2011.

\bibitem{viswanathan2011physics}
G.M. Viswanathan, M.G.E. da~Luz, E.P. Raposo, and H.E. Stanley.
\newblock {\em The Physics of Foraging: An Introduction to Random Searches and
  Biological Encounters}.
\newblock Cambridge Univ Pr, 2011.

\bibitem{vitter1985random}
Jeffrey~S Vitter.
\newblock Random sampling with a reservoir.
\newblock {\em ACM Transactions on Mathematical Software (TOMS)}, 11(1):37--57,
  1985.

\bibitem{weng2012competition}
L~Weng, A~Flammini, A~Vespignani, and F~Menczer.
\newblock Competition among memes in a world with limited attention.
\newblock {\em Scientific Reports}, 2, 2012.

\bibitem{Lilian2013srep}
L.~Weng, F.~Menczer, and Y.-Y. Ahn.
\newblock Virality prediction and community structure in social networks.
\newblock {\em Sci. Rep.}, 3(2522), 2013.

\bibitem{weng2013role}
Lilian Weng, Jacob Ratkiewicz, Nicola Perra, Bruno Gon{\c{c}}alves, Carlos
  Castillo, Francesco Bonchi, Rossano Schifanella, Filippo Menczer, and
  Alessandro Flammini.
\newblock The role of information diffusion in the evolution of social
  networks.
\newblock {\em arXiv preprint arXiv:1302.6276}, 2013.

\end{thebibliography}

\end{document}